\documentclass[multphys,vecphys]{svmult}

\usepackage{makeidx}         
\usepackage{graphicx}        
\usepackage{multicol}        
\usepackage[bottom]{footmisc}

\makeindex             

\begin{document}

\title*{AGN Heating of Cooling Flow
  Clusters: Problems with 3D Hydrodynamic Models}
\titlerunning{Problems with Heating Cooling Flows}
\author{John C. Vernaleo\inst{1,2}\and
Christopher Reynolds\inst{1,3}}
\authorrunning{Vernaleo and Reynolds}
\institute{Department of Astronomy, University of Maryland, College
  Park, MD 20742 \and
\texttt{vernaleo@astro.umd.edu}
\and \texttt{chris@astro.umd.edu}}
%
%
\maketitle

\section{Introduction}
\label{sec:intro}
Relaxed galaxy clusters have central cooling times less than
the age of the cluster.  However, there are observational limits to
the amounts of cool gas 
present, and XMM-Newton spectroscopy shows nothing
below 
$\sim\frac{1}{3}T_{virial}$ ($1 - 2 keV$).  This discrepancy is the heart
of the classical cooling flow problem.

Viewed from another angle, 
the galaxy luminosity function is truncated at the high
end~\cite{2003ApJ...599...38B}.
So whatever offsets cooling probably also stops the formation of
massive galaxies.
This must occur on many mass and temperature scales:
therefore some self regulation seems to be required.
AGN (Active Galactic Nuclei) are often given as a possible solution.
AGN inject energy on the same order as cooling
luminosity ($\sim10^{45} - 10^{46} \mbox{ erg } s^{-1}$).
They are fed by accretion, so self regulation may come naturally.

Numerous observations indicate that AGN can have 
an impact on large scale structure. 
We have performed a set of high 
resolution, three dimensional simulations of a jetted AGN embedded in a 
relaxed cooling cluster~\cite{2006ApJ...645...83V}. To the best of our
knowledge, these are the first simulation to include both full
jet dynamics and a feedback model.  These ideal
hydrodynamic simulations show that,
although there is enough energy present to offset cooling on average, the 
jet heating is not spatially deposited in a way that can prevent 
catastrophic cooling of the cluster.

\section{Models}
\label{sec:models}

For comparison, we start with a pure cooling cluster.  All other
clusters start with the same, with some additional
effects added.  The initial cluster is modelled after a rich, relaxed
cluster with a 
$\beta$-law density profile:  $r_{core}=100\,{\rm kpc}$,
$n_0=0.01\,{\rm cm}^{-3}$, and
$c_{s}=1000\,{\rm km}\,{\rm s}^{-1}$.  The cooling is modelled as thermal
bremsstrahlung emission, following~\cite{2002ApJ...581..223R}.
Due to $n^2$ dependence of the ICM cooling, in the absence of any
feedback, cooling runs away, showing a featureless increase with time
It gets to a level we set as 'catastrophic' by around 250 Myrs.

We added several types of feedback to our models.  The first type is a
single jet outburst lasting 50 Myrs.
To model
actual feedback, the velocity of the jet (and hence the kinetic
luminosity of the source) was varied based on $\dot{M}$ across the
inner edge of the simulation.
This was
done with various time delays (up to 100 Myrs) and efficiencies ($\eta=0.00001-0.1$).

\begin{figure}
  \centering
  \includegraphics[height=3.5cm]{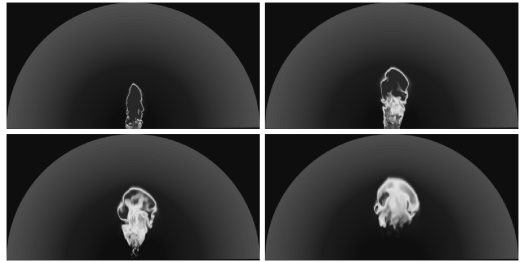}
  \caption{Entropy for single jet simulation.}
  \label{fig:jet}
\end{figure}

For the single burst jet, with a kinetic luminosity of
$L_{kin}=9.3\times10^{45}\mbox{ erg }
s^{-1}$ and a Mach $10.5$ jet (see
also~\cite{2002MNRAS.332..271R}), catastrophic cooling was delayed by
about 50 Myrs.  The results of this simulation (which show the bubble
the jet inflates) can be seen in Figure~\ref{fig:jet}.

For feedback, mass flow across inner boundary was calculated and used
to set a jet velocity assuming some efficiency $\eta$ of the central
blackhole using the formula
$v_{jet}=(\frac{2\eta\dot{M}c^2}{A\rho})^{\frac{1}{3}}$.

\begin{figure}
  \parbox{5.0cm}{
  \centering
  \includegraphics[height=3.2cm]{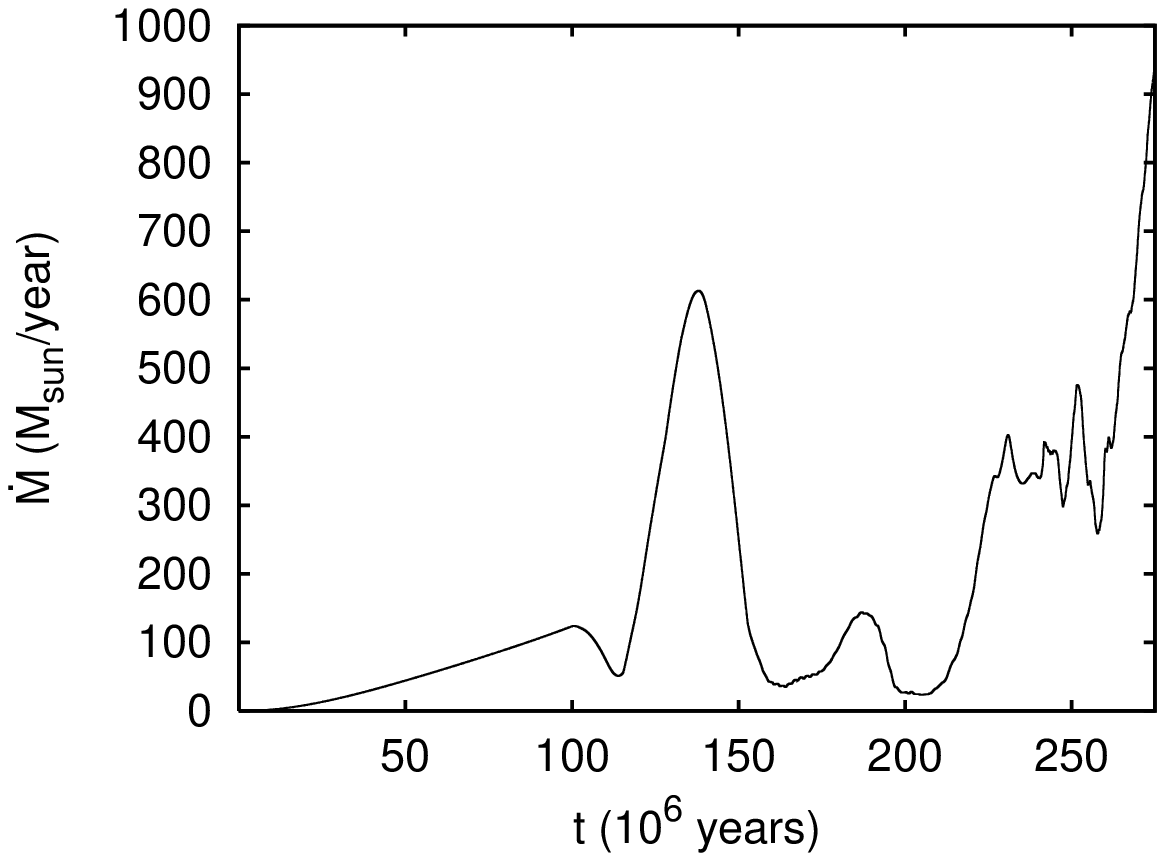}
  \caption{Mass accretion for delayed feedback.}
  \label{fig:mdotI}
}
\hspace{0.5cm}
\parbox{5.0cm}{
  \centering
  \includegraphics[height=3.2cm]{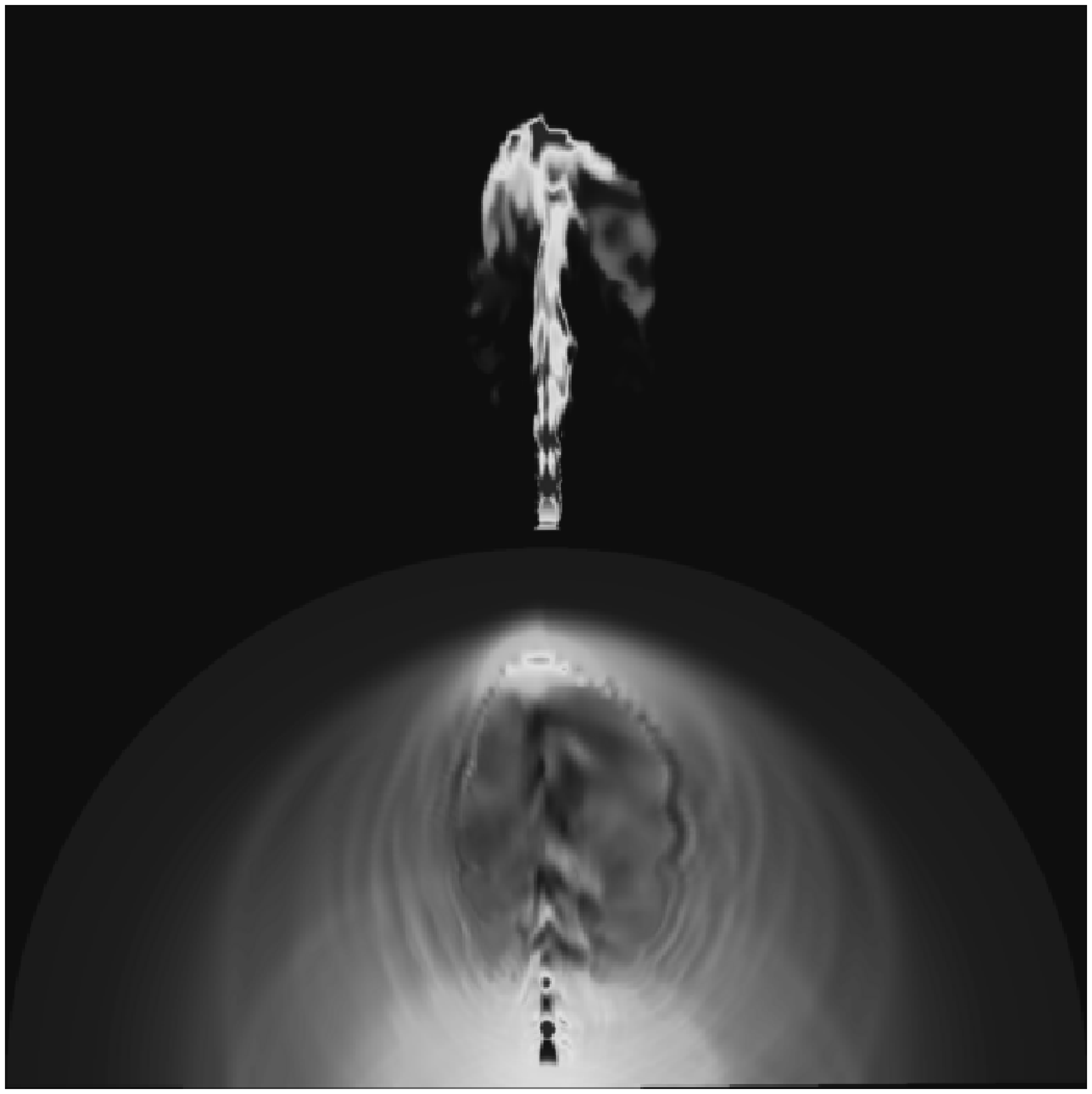}
  \caption{Channel formation (Temperature (top)
        Pressure (bottom)).}
  \label{fig:channel}
}
\end{figure}

The most realistic model seems to be the low efficiency
($\eta=10^{-4}$) model 
with a delay of 100 Myrs (close to the dynamical time for the
galaxy).  Even this only delays the cooling catastrophe (see
Figure~\ref{fig:mdotI} for mass accretion rates).

The jet seems to cut a channel (see Figure~\ref{fig:channel}) in the
ICM which allows it to avoid 
heating the inner regions.  This explains why simulations with
bubbles placed in the center can do better at halting cooling than
jets, but are less realistic.

\section{Conclusions}
\label{sec:conc}
We have preformed the first simulations that we are aware of to
include both the full dynamics of a jet and an feedback model.  When the
full dynamics of the jet are included, ideal hydrodynamics
interactions do not seem able to offset cooling on average, even
though they are energetically
capable of doing so.
We conclude that either some 
physical process beyond that captured by our ideal hydrodynamic 
simulations (e.g., plasma transport processes, cosmic ray heating, 
dramatic jet precession, or ICM turbulence) is relevant for thermalizing 
the AGN energy output, or the role of AGN heating of cluster gas has been 
overestimated.

\section{ZEUS-MP}
\label{sec:zeus}

All simulations were done using the ZEUS-MP 3D parallel
hydrocode (a version of the code
in~\cite{1992ApJS...80..753S}).  We
have updated and modified the NCSA release (v1.0).  Our modifications
(v1.5) and documentation are publicly available at:
\texttt{http://www.astro.umd.edu/\char'176 vernaleo/zeusmp.html}
There is also a version 2 of ZEUS-MP which we are not affiliated with.
\section{Acknowledgments}
\label{sec:ack}

Simulations
were performed on the Beowulf cluster (``The
Borg'') supported by the Center for Theory and Computation (CTC) in
the Department of Astronomy, University of Maryland.  This work was 
partly funded by the {\it Chandra} Cycle-5 Theory \& Modelling program 
under grant TM4-5007X.

%
%
\bibliographystyle{plain}
\bibliography{proc}
%


\printindex
\end{document}